\documentclass[aps,pre,reprint,groupedaddress, superscriptaddress]{revtex4-1}
 
\usepackage{amssymb}

\usepackage{graphicx}
\usepackage{lipsum}

\usepackage{makecell}
\usepackage{amsmath}
\begin{document}
\title{Pixel-wise classification in graphene-detection with tree-based machine learning algorithms}
\author{Woon Hyung Cho}
%\email[cuh1996@uos.ac.kr]{}
\affiliation{Department of Physics, University of Seoul, Seoul 02504, Korea}
\affiliation{Department of Smart Cities, University of Seoul, Seoul 02504, Korea}

\author{Jiseon Shin}
%\email[jsshin9121@uos.ac.kr]{}
\affiliation{Department of Physics, University of Seoul, Seoul 02504, Korea}

\author{Young Duck Kim}
%\email[ydk@khu.ac.kr]{}
%\affiliation{Department of Physics, Kyung Hee University, Seoul 02447, Korea}
\affiliation{Department of Physics, Department of Information Display, KHU-KIST Department of Converging Science and Technology, Kyung Hee University, Seoul 02447, Korea}

\author{George J. Jung}
\email[gjtbgf@g.uos.ac.kr]{}
%\email[jeiljung@uos.ac.kr]{}
\affiliation{Department of Physics, University of Seoul, Seoul 02504, Korea}
\affiliation{Department of Smart Cities, University of Seoul, Seoul 02504, Korea}

\begin{abstract}
Mechanical exfoliation of graphene and its identification by optical inspection is one of the milestones in condensed matter physics that sparked the field of 2D materials. Finding regions of interest from the entire sample space and identification of layer number is a routine task potentially amenable to automatization. 
We propose supervised pixel-wise classification methods showing a high performance even with a small number of training image datasets that require short computational time without GPU. We introduce four different tree-based machine learning algorithms -- decision tree, random forest, extreme gradient boost, and light gradient boosting machine. We train them with five optical microscopy images of graphene, and evaluate their performances with multiple metrics and indices. 
We also discuss combinatorial machine learning models between the three single classifiers and assess their performances in identification and reliability. The code developed in this paper is open to the public and will be released at \url{github.com/gjung-group/Graphene_segmentation}.
\end{abstract}

\maketitle
\section{Introduction}

The fact that a few layers of graphene sheet can be prepared
by simple mechanical exfoliation ~\cite{novoselov2004electric, yi2015review} 
has facilitated a rapid growth of graphene and other two-dimensional (2D) van der Waals (vdW) materials research. 
In particular, graphene has been studied in a wide range of applications in recent years due to its unique electrical, mechanical, and optical properties~ \cite{neto2009electronic,koppens2014photodetectors,geim2007rise, lee2008measurement,cao2020elastic,bonaccorso2010graphene, geim2013van, ajayan2016van}. 
The recent discovery of a robust unconventional superconductivity in twisted graphene systems \cite{bistritzer2011moire, cao2018unconventional, cao2018correlated, park2021tunable, hao2021electric} has reinvigorated research in graphene and other two-dimensional (2D) van der Waals (vdW) materials~\cite{chhowalla2013chemistry, novoselov2005two,zeng2010white,watanabe2004direct,reich2014phosphorene,chen2017rising}.

%xu2014spin, xiao2012coupled,jones2013optical,splendiani2010emerging,wang2020correlated}. 

%
After the mechanical exfoliation of 2D materials, the number of layers of graphene or other vdW materials can be identified by various techniques including atomic force microscopy (AFM)~\cite{huang2015reliable,shearer2016accurate}, Raman spectroscopy~\cite{saito2011raman,huang2015reliable}, or optical microscopy (OM)~\cite{blake2007making, li2013rapid, jessen2018quantitative, huang2019optical}. 
Among them, the most commonly used method is based on the optical contrast between single and multilayer graphene layers with different thicknesses in RGB color space of OM
images of materials placed on a substrate of specific thickness ~\cite{blake2007making, li2013rapid,jessen2018quantitative, huang2019optical}.
However, it is a time consuming process to process more than $\sim 10^3$ scanned OM images to identify the interesting few layers exfoliated flakes regions deposited on the substrate. Here we propose a practical machine learning image recognition method that  can be used to quickly identify specific target few layers graphene regions.

Traditionally, machine learning (ML) based techniques have been applied succesfully in many different fields of industrial applications or services requiring repetitive human labor. Especially, image classification using deep learning (DL) has emerged as a game changer technique that has allowed drastic reduction of analysis time from hours to seconds.
For example, the convolutional neural network (CNN) has been used in biomedical fields, such as abdominal CT scan, cell, hippocampus, and pancreas segmentations~\cite{ronneberger2015u, oktay2018attention,vizcaino2021pixel}, and in analyzing big image data obtained from satellites~\cite{zhao2018building, leon2020big} providing significant aid to error-prone human eyes.

Recent works have used ML techniques to identify the number of layers in a thin film of materials. These researches have employed supervised learning such as support vector machines (SVM)~\cite{lin2018intelligent, yang2020automated}, deep neural network (DNN)~\cite{masubuchi2019classifying, greplova2020fully, shin2021fast}, U-Net which belongs to CNN~\cite{saito2019deep, dong20213d, siao2021machine}, or unsupervised learning such as clustering \cite{masubuchi2019classifying}. The image classifications in the earlier works, however, require too many images for the training dataset that need to be labeled accordingly with layer number, for example, $\sim 10^3-10^5$ labeled images for DNN~\cite{masubuchi2019classifying, greplova2020fully, shin2021fast}, and $\sim 10^2$~\cite{siao2021machine,dong20213d} or $10^3$ labeled images for U-Net which are augmented from less than 50 labeled images~\cite{saito2019deep}, and more than a dozen GB of GPU memory to process a batch of image data.

\begin{figure*}[t]
\begin{center}
\includegraphics[width=1\textwidth]{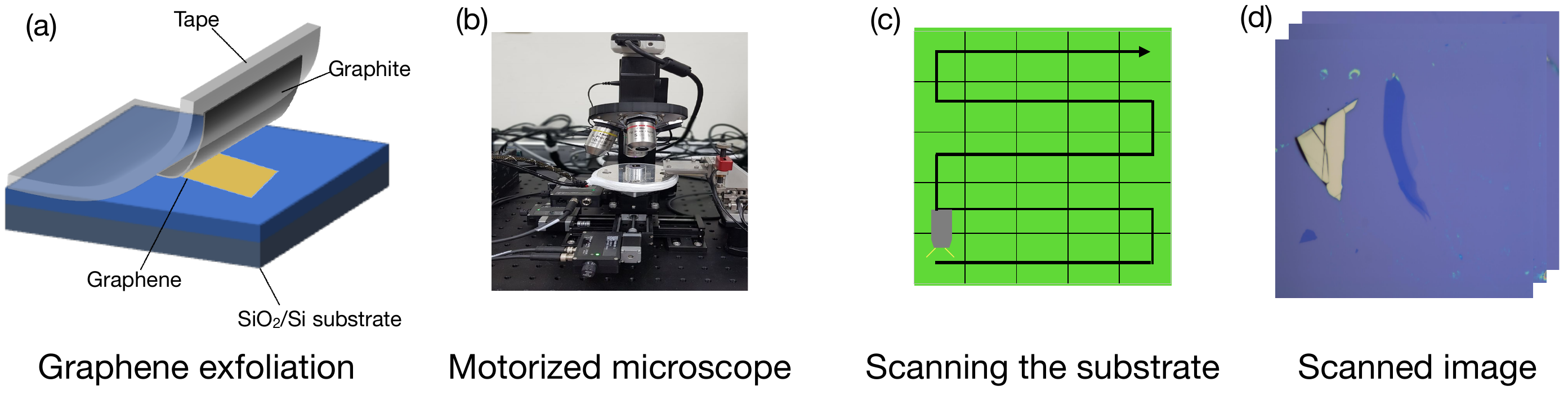}
\end{center}
\caption{
(a) Schematic diagram for exfoliation of graphene from graphite using the scotch tape on top of the substrate SiO$_2$/Si. (b) Motorized microscope for scanning the sample. (c) Schematic diagram for the entire sample is denoted by the green panel, the path indicated by the black arrow for scanning the sample to find the graphene, and the camera of the motorized microscope illustrated by the gray square. (d) Scanned images by the motorized microscope along with the black arrows in (c). 
}
\label{fig1}
\end{figure*}

%
%To lower such entry barriers, 
In this paper we suggest a handy classification tool for detecting a specific number of graphene layers pixel by pixel using supervised ML algorithms requiring only a few OM images for training. We compare the performance of four different tree-based ML algorithms such as decision tree (DT), random forest (RF), extreme gradient boosting (XGBoost), and light gradient boosting machine (LightGBM) in terms of accuracy, precision, recall, F1 score, and indices such as receiver operating characteristic (ROC) curve, area under curve (AUC), intersection over union (IOU). We take account of several combinations between RF, XGB, and LGBM entailing improvements in multiple metrics compared to using a single algorithm. Our source code was built with scikit-learn~\cite{pedregosa2011scikit} ML library and is provided as open-source (See Ref.~\cite{source}).

This paper is organized as follows. In Sec.~\ref{sec2} we first describe the sample preparation for OM images, feature extraction using several filters to pre-treat the images for ML algorithms. We elaborate on the four different tree-based ML algorithms that we employed in this work in Sec.~\ref{sec3}. In Sec.~\ref{sec4}, we evaluate and discuss the performances of a single tree-based ML algorithm and combinations for fusions of them using the several metrics. In Sec.~\ref{sec5} we present the conclusions.

\section{Data Preprocessing}\label{sec2}
In the following we describe the data preprocessing techniques used for this work that consists in dataset preparation from the OM images obtained from the experimental system and feature extraction from the images by using different filters. 

\subsection{Dataset preparation}

The graphene samples in the present work were exfoliated using scotch tapes and placed on top of 285~nm SiO$_2$/Si substrates as shown in Fig.~\ref{fig1} (a). We use a motorized microscope as seen in Fig.~\ref{fig1} (b) to scan the $20\times$ magnified images with $1832\times1372$ resolution. We scan the entire sample along the path indicated by the black arrow as shown in Fig.~\ref{fig1} (c). Afterwards, we get scanned OM images like Fig.~\ref{fig1} (d). To improve the speed of image preprocessing and classification, images were resized with $458\times343$ pixels which are moderately large for finding graphene. We only use 5 OM images for training and 2 images for testing. However, in terms of the number of segmentation resolved by pixel, we use in total 19,636,750 pixels for the training dataset and 785,470 pixels for the testing dataset.
The mask for labeling was made manually using the APEER platform~\cite{apeer}.

\begin{figure*}[t]
\begin{center}
\includegraphics[width=0.8\textwidth]{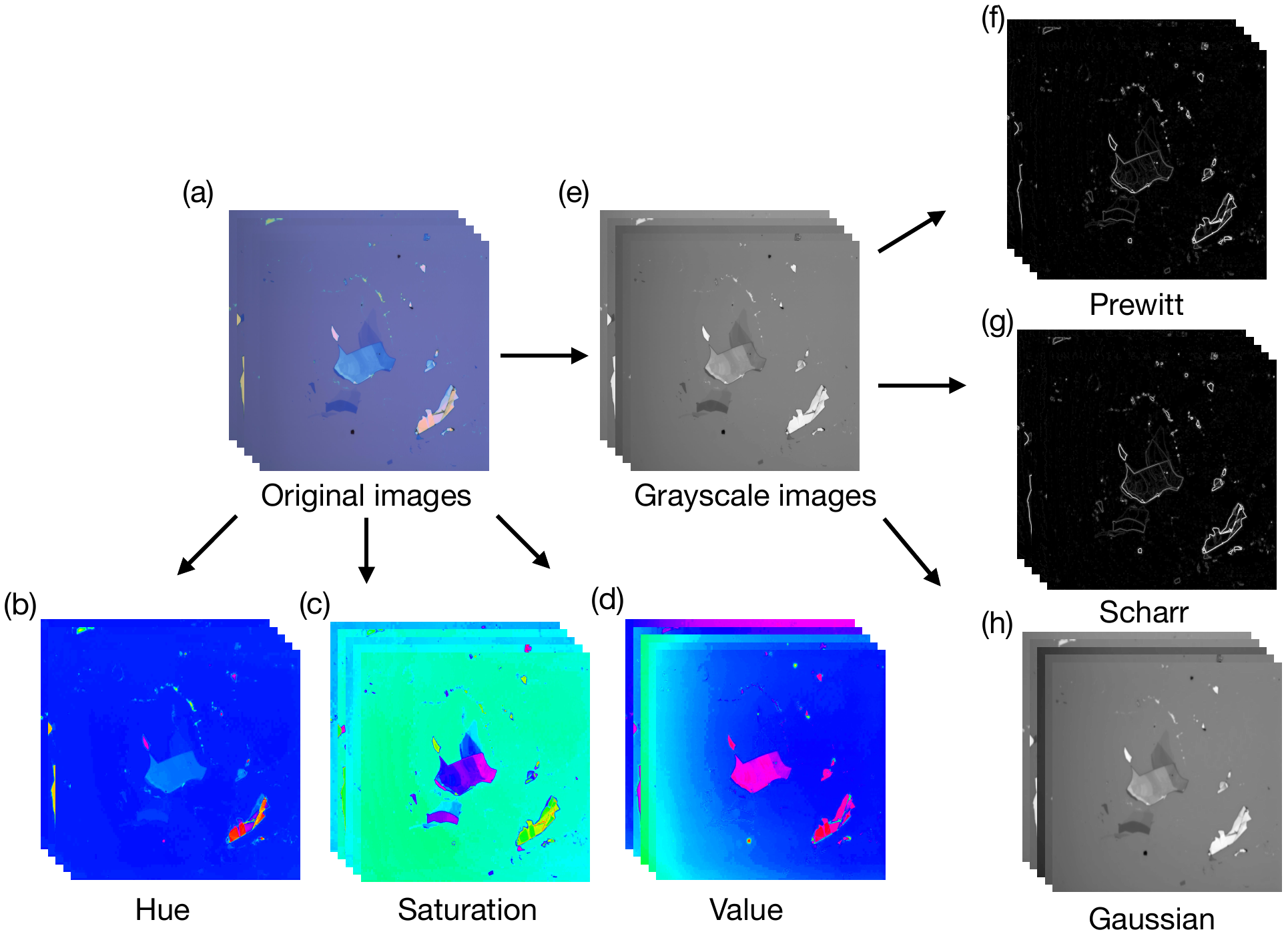}
\end{center}
\caption{
Flowchart for data preprocessing for ML feature extraction. (a) Original images from motorized spectroscopy. (b) Map for the hue (H) component of the original images. (c) Map for the saturation (S) component of the original images. (d) Map for the value (V) component of the original images. (e) Images in grayscale. (f) Resultant map of the Prewitt edge filter for the grayscale images. (g) Resultant map of the Scharr edge filter. (h) Resultant map of the Gaussian filter.}
\label{ML_flow}
\end{figure*}

\subsection{Feature Extraction}

Feature extraction is one of the most important key elements in ML. Here we use the HSV color space and several commonly used filters that help us extract relevant features to recognize and classify the monolayer graphene from the background as we describe in the following.
%\subsubsection{Color features}
%
There are several color bases for generating arbitrary colors that we are familiar with, such as RGB (red, green, blue) and CMYK (cyan, magenta, yellow, black). However, here we introduce the HSV (hue, saturation, value) color space, which is a familiar way for humans to perceive color. The HSV color space describes colors with their hue (H) together with saturation (S), namely, amount of gray, and value (V) of brightness or luminous intensity as illustrated in Fig.~\ref{ML_flow} (b-d). Each pixel of the graphene images obtained from the motorized microscope which were originally expressed in RGB color space are transformed to HSV~\cite{wen2004color, chen2007identifying}.

%\subsubsection{Edge features using filters}
%
We use several filters, the Prewitt, Scharr, and Gaussian filters in order to extract the relevant features. We use the grayscale images for this process as shown in Fig.~\ref{ML_flow} (e) for the following two reasons. Firstly, the grayscale images preserve the essential information such as edge, shape, and texture information from their original RGB representation. Secondly, we can reduce complexity and unnecessary computational cost~\cite{bui2016using}. An edge of any object in images can be defined as a boundary where the value of brightness changes discontinuously, leading to a very sharp change in the associated intensity gradient. Hence, we can find abrupt changes in the intensity values for each pixel over the entire image given in HSV space, and find spots where its derivative is maximum. We utilize the $3 \times 3$ Prewitt, and Scharr operators for the $x$ (vertical) and $y$ (horizontal) direction as defined below \cite{prewitt1970object, jahne1999handbook} 

\begin{displaymath}
\textrm{Prewitt}_x =\frac{1} {3} \begin{pmatrix}
1& 0 & -1\\
1 & 0 & -1\\
1& 0 & -1
\end{pmatrix}, 
\end{displaymath}

\begin{equation}
\textrm{Prewitt}_y =\frac{1} {3} \begin{pmatrix}
1& 1& 1\\
0 & 0 & 0\\
-1 & -1 & -1
\end{pmatrix},
\end{equation}

and
\begin{displaymath}
\textrm{Scharr}_x = \frac{1} {32}\begin{pmatrix}
3& 0 & -3 \\
10 & 0 & -10\\
3& 0 &-3
\end{pmatrix},
\end{displaymath}

\begin{equation}
\textrm{Scharr}_y = \frac{1} {32}\begin{pmatrix}
3& 10 & 3 \\
0 & 0 & 0\\
-3& -10 &-3
\end{pmatrix}.
\end{equation}
As an example, we present the resultant map for an image processed through the Prewitt and Scharr filters in Fig.~\ref{ML_flow} (f) and (g), respectively. After the convolution of the two operators with our grayscale image, we get resultant $F_x$ and $F_y$ matrices of the same size as our source image. Hence, the magnitude for each pixel is given by
\begin{equation}
\textrm{Magnitude} = \sqrt{ F_x^2 + F_y^2 }.
\label{Magn}
\end{equation}
Afterwards, we use the Gaussian smoothing kernel based on the 2D Gaussian function of Eq.~(\ref{gaus}). 

\begin{equation}
G(x,y)= \frac{1} {2\pi\sigma^2}e^{-\frac{x^2+y^2}{2\sigma^2}}
\label{gaus}
\end{equation}
The Gaussian kernel~\cite{marr1980theory}  raises the quality of our model by blurring redundant noises as illustrated in Fig.~\ref{ML_flow} (h). For example, the $5\times5$ Gaussian kernel for $\sigma=1$ is  

\begin{equation}
G^{\sigma = 1}_{5\times5} =
\frac{1} {273} \begin{pmatrix} 
1& 4 & 7 &4&1 \\
4& 16& 26 &16&4 \\
7& 26& 41&26&7\\
4& 16 & 26 &16&4 \\
1& 4 & 7 &4&1
\end{pmatrix}.
\end{equation}

\begin{figure*}[t]
\begin{center}
\includegraphics[width=0.9\textwidth]{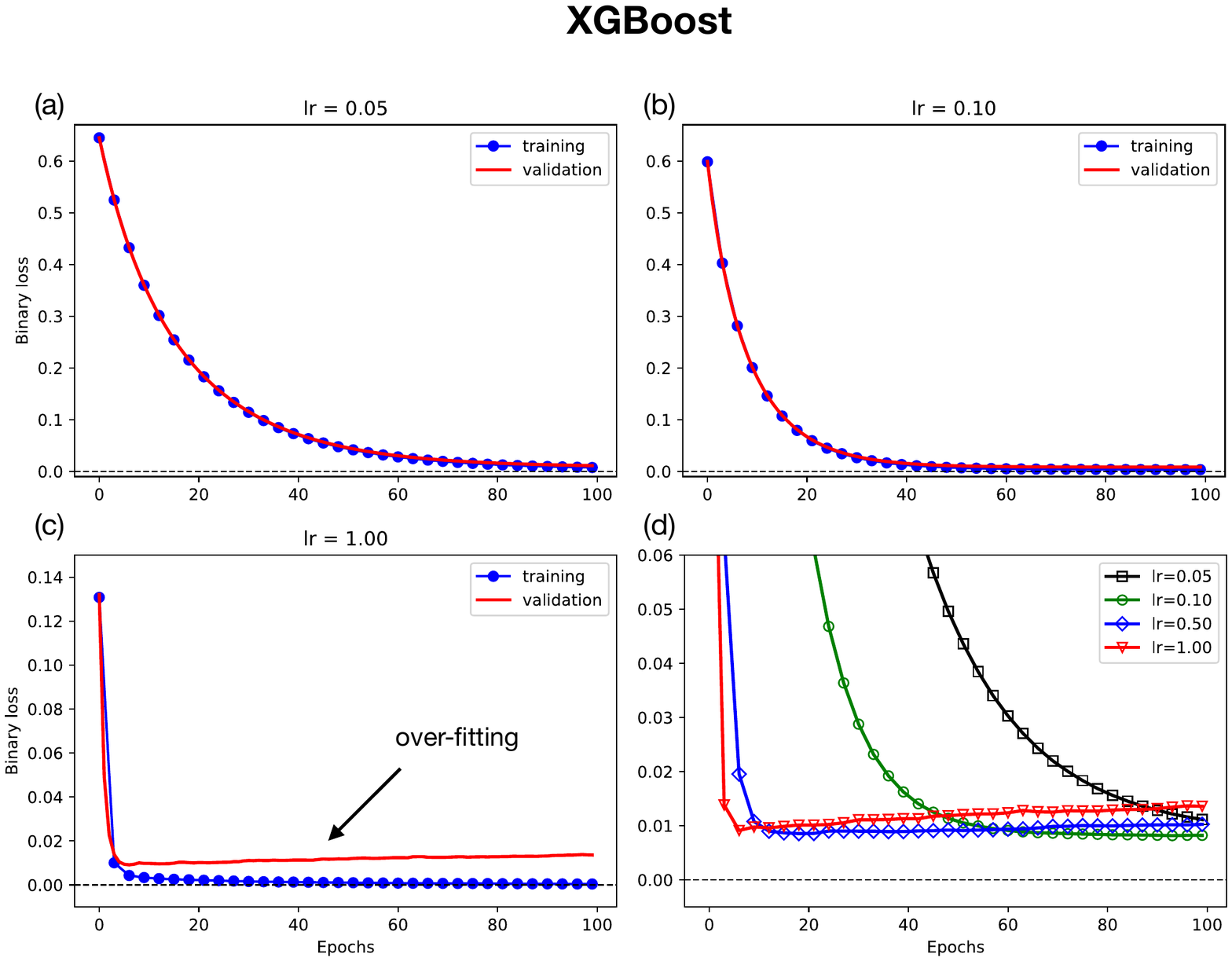}
\end{center}
\caption{
Training (blue filled circles) and validation (red solid line) loss in XGB algorithm using the binary cross-entropy loss function for the learning rates (a) lr = $0.05$, (b) $0.1$, and (c) $1.0$. (d) Validation losses for four different learning rates, lr = $0.05$ (black squares), $0.10$ (green circles), $0.5$ (blue diamonds), and $1.0$ (red triangles). XGB is over-fitted with the training dataset when the learning rate is larger than lr $\approx 0.5$.}
\label{loss_xgb}
\end{figure*}

\section{Theoretical frameworks}\label{sec3}
In the following, we briefly summarize the theoretical footing of the four tree-based ML classification algorithms employed in the current work. The simplicity and efficiency of these models are at the heart of the practical applicability our model to reduce the number of training images and shorten the classification time. This section is devoted mostly to assess the training and validation of the more advanced extreme gradient boost and light gradient boosting machine models. 

\subsection{Decision tree}
The decision tree (DT) is one of the basic algorithms in ML, 
that as its name indicates infers the predicted label following a tree-like decision framework. 
Firstly, we have a root (a single leaf) to begin with and we assign a label to this root according to a majority vote among all the labels over the training set. For example, we can imagine that the label could be the length of the petal in the famous classification problem for Iris flowers. Then, we can split the root into two groups depending on whether or not the dataset satisfies the label and evaluate the effect of splitting over the iterations by calculating a measure which is called gain. The gain quantifies the improvement of the performance of our model due to the splitting. Among the possible splits, we can either choose the one that maximizes the gain or choose not to split the leaf. The merit of this algorithm is that we can see the procedures of decision-making in the algorithm. However, the major disadvantage of DT is the high risk of over-fitting to the training set which can be caused by an outlier that is chosen to be a single node even though it is non-representative~\cite{quinlan1996learning,myles2004introduction,loh2011classification}.

\begin{figure*}[t]
\begin{center}
\includegraphics[width=0.9\textwidth]{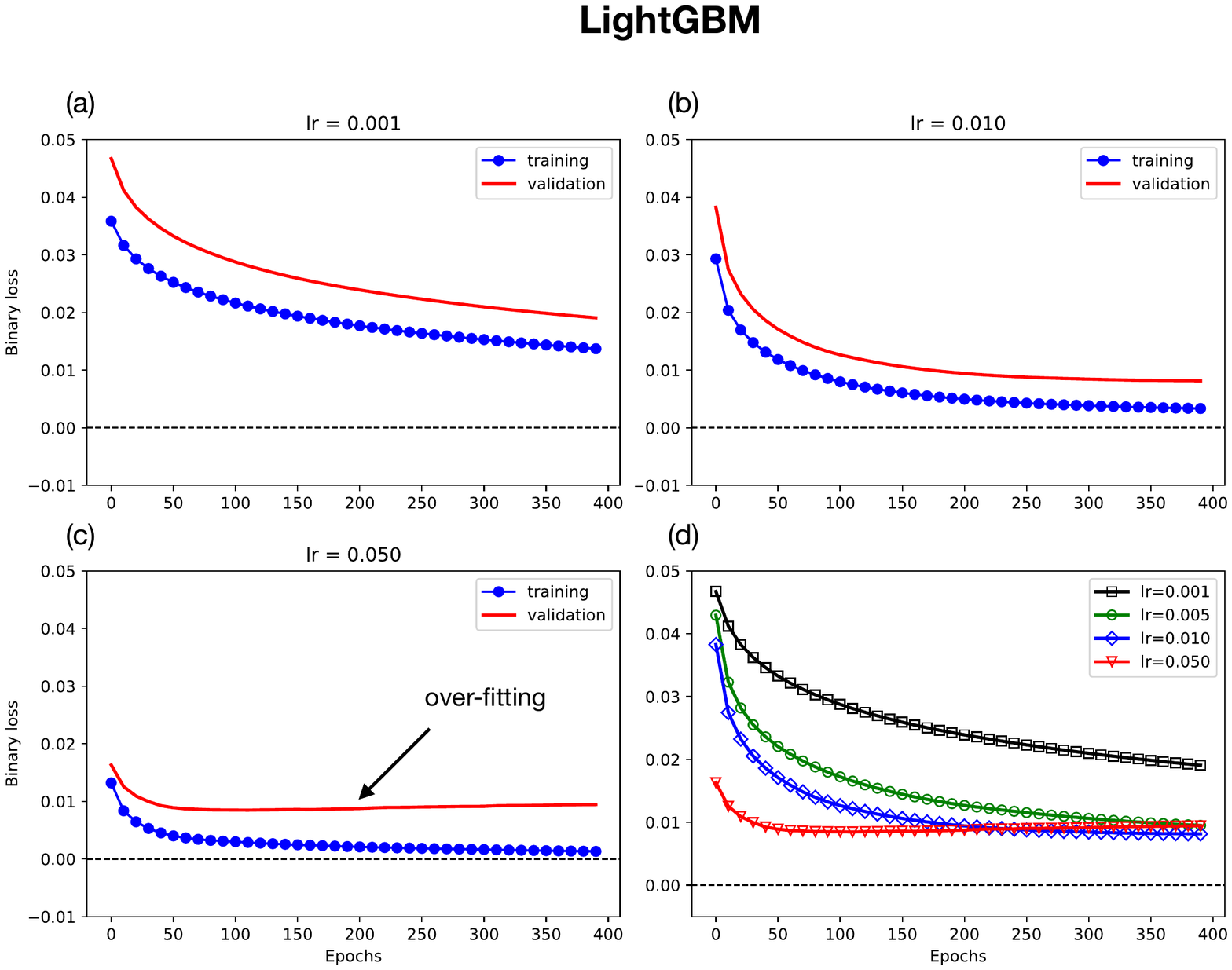}
\end{center}
\caption{
Training (blue filled circles) and validation (red solid line) loss in LGBM algorithm using the binary cross-entropy loss function for the learning rates (a) lr = $0.001$, (b) $0.01$, and (c) $0.05$. (d) Validation losses for four different learning rates, lr = $0.001$ (black squares), $0.005$ (green circles), $0.01$ (blue diamonds), and $0.05$ (red triangles). LGBM is over-fitted with the training dataset as the learning rate increases more than lr $\approx 0.05$.}
\label{loss_lgbm}
\end{figure*}

\subsection{Random forest}
The random forest (RF) is literally an ensemble of the decision trees with randomness such that RF generates multiple decision trees that chooses a class by majority vote among the trees and aggregates them by using their average for a regression. Therefore, RF is less vulnerable to over-fitting than DT. However, the decision-making could slow down as the size of the dataset increases~\cite{breiman2001random,liu2012new,akar2012classification}.

\subsection{XGBoost}
The extreme gradient boost (XGB or XGBoost) is a scalable end-to-end ML algorithm which was proposed by Chen and Guestrin in 2016~\cite{chen2015xgboost,chen2016xgboost}. XGB is based on a gradient boosting decision tree (GBDT). GBDT combines base-learners (e.g., DT) into a single strong learner over the many iterations by fitting the base-learners using a loss function (e.g., the mean squared error). The aim of gradient boosting is to train the model to minimize the loss function using the functional gradient descent which leads the next iteration toward the direction of negative gradient. While GBDT uses only the first-order derivatives of the loss function, XGB utilizes the Newton-Raphson method for the functional gradient. Namely, the second-order derivatives of the loss function are used in the fitting procedure. Consequently, XGB proposes a newly distributed algorithm for tree searching, outperforming RF in general but XGB has the drawback that it is generally more time-consuming to perform~\cite{friedman2017elements}. 

Fig.~\ref{loss_xgb} shows the training and validation losses in XGB using the binary cross-entropy loss function 
% {\bf 
\begin{equation}
L_{\log} %= -\log \operatorname{Pr}(y|p) 
= \frac{1}{N} \sum_{i = 0}^{N-1} -(y_i \log (p_i) + (1 - y_i) \log (1 - p_i))
\end{equation}
%  }
as implemented in scikit-learn~\cite{pedregosa2011scikit} as a 
function of the epochs for the three different learning rates (lr)
where $y \in \{0,1\}$ is the true label and $p = \operatorname{Pr}(y = 1)$ is the probability and the $i$ index runs over all sample points. 
The training loss decreases and approaches zero for all cases as the epoch increases. For lr~$= 0.05$ and $0.10$ the validation loss decreases monotonically, while the validation loss has a local minimum at the epoch~$=16$ and grows as the epoch increases for lr~$=1.0$, implying that the model is over-fitted with the training dataset. Our model starts over-fitting when lr is larger than $\approx 0.5$ as shown in Fig.~\ref{loss_xgb} (d). Thus, in this paper, we use lr $=0.1$ for XGB unless stated otherwise.

\subsection{LightGBM}
The light gradient boosting machine (LGBM or LightGBM) also originates from GBDT and inherits many strengths from XGB. However, LGBM has two significant technical differences from XGB in structuring trees which are called gradient-based one side sampling (GOSS) and exclusive feature bundling (EFB)~\cite{ke2017lightgbm}. 

The GOSS is a sampling tool that keeps the instances with large gradients but randomly drops out some portion of the ones with small gradients. In EFB two sparse features which are nearly exclusive integrate into one to reduce the number of instances. In short, GOSS and EFB reduce the number and the size of data instances remarkably while at the same time keeping high accuracy. 
Consequently, LGBM is faster and requires less memory, as its name suggests ‘‘light". However, LGBM is prone to over-fitting when the size of the dataset is small~\cite{zhao2019predicting}. 

\begin{figure*}[t]
\begin{center}
\includegraphics[width=1\textwidth]{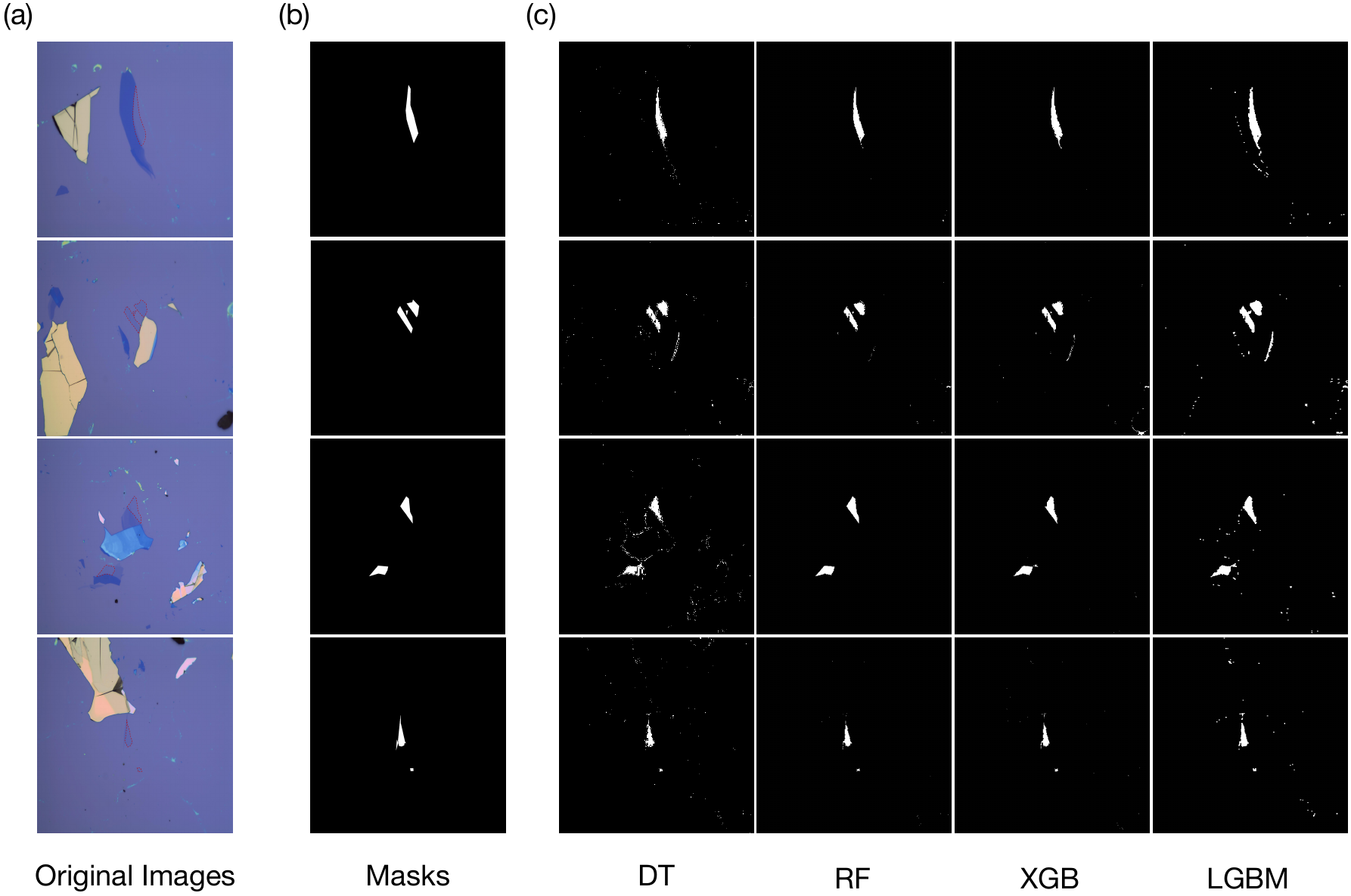}
\end{center}
\caption{
(a) Source images of the domain of interest. The flakes of monolayer graphene are enclosed by red dashed lines. (b) The flakes of monolayer graphene are labeled as 1 (white) and the rest is labeled as 0 (black). (c) Resultant images of pixel-wise classifications for the four algorithms, DT, RF,  XGBoost (XGB), and LightGBM (LGBM).
}
\label{result}
\end{figure*}

We show the training and validation losses for different learning rates lr~$=0.001$, $0.01$, and $0.05$ in Fig.~\ref{loss_lgbm} (a-c) using the binary cross-entropy loss function. One can see that LGBM model for our graphene OM image datasets requires to be stopped early at the epoch $=20$ to avoid over-fitting where the validation loss starts increasing when lr~$=0.05$. Compared to other learning rates, the loss starts increasing upon the epochs when lr~$\gtrsim 0.05$ as shown in \ref{loss_lgbm} (d). Therefore, we choose lr $=0.005$ for LGBM throughout this paper.

\section{Results and Discussions}\label{sec4}

\subsection{Single classifiers}
We train the ML models for each pixel in an OM image with each one of the four classification models and 
%{\bf 
use the five-fold stratified cross-validation to maintain objectivity~\cite{kohavi1995study}. 
Thus, we obtain the probability of having a monolayer graphene sheet at each pixel, and round off the probability from the first decimal place so that we assign $0$ to pixels predicted as background, and $1$ to the ones predicted as monolayer graphene at each pixel. 
In Fig.~\ref{result} we visualize our results for the four tree-based ML algorithms together with the original OM images and the target images labeled for the monolayer graphene. The graphene monolayers in the OM images in Fig.~\ref{result} (a) are enclosed by the red dashed line. As shown in Fig.~\ref{result} (b), the pixels for monolayers are labeled as 1 (white) and the others are labeled as 0 (black). Fig.~\ref{result} (c) shows the classification results from DT, RF, XGB, and LGBM algorithms, respectively. Each algorithm appears to well differentiate the monolayer graphene from the OM image. 
%
%{\bf As our dataset consists of either a monolayer of graphene pixels or background pixels which leads to an imbalance in detecting monolayer graphene correctly, introduce several evaluation indices as follows.}
Since background pixels are dominant, it leads to an highly imbalanced pool of pixels. Therefore we introduce several metrics and indices to evaluate performance of the model
\begin{table*}[]
\begin{tabular}{|c|c|c|c|c|}
\hline
\diaghead(-2,1){aaaaaaaaaaaaaa}%
{Algorithm}{metric}     
              & \multicolumn{1}{l|}{Accuracy (\%)} & \multicolumn{1}{l|}{Precision (\%)} & \multicolumn{1}{l|}{Recall (\%)} & \multicolumn{1}{l|}{F1 Score (\%)} \\ \hline
DT & 99.68                             & 73.70                              & 77.92                            & 75.36                             \\ \hline
RF & 99.79                             & 89.61                             & 76.54                            & 82.09                             \\ \hline
XGB       & 99.72                             & 88.01                              & 75.26                            & 80.80                             \\ \hline
LGBM      & 99.77                             & 91.23                              & 73.37                            & 80.96                             \\ \hline
\end{tabular}
\caption{Evaluations of the four different tree-based classification algorithms, decision tree (DT), random forest (RF), XGBoost (XGB), and lightGBM (LGBM) using the four metrics such as accuracy, precision, recall, and F1 score defined in Eqs.~(\ref{eqAccuracy}-\ref{eqF1}).}
\label{table_metric}
\end{table*}

First, the accuracy as defined in Eq. (\ref{eqAccuracy}) is the most basic metric to evaluate the performance of a classification ML model.  

\begin{equation}
\textrm{Accuracy} = \frac{\textrm{TP + TN}} {\textrm{TP + FN + TN + FP}},
\label{eqAccuracy}
\end{equation}
where TP (TN) stands for the case where our model correctly predicts a graphene (background) pixel, while FP (FN) represents the opposite case where our model incorrectly predicts a background (graphene) pixel as a graphene (background) pixel.
The precision which is defined as Eq. (\ref{eqPrecision}) is obtained by evaluating the ratio of correctly predicted graphene pixels.
%true graphene pixels to pixels predicted as graphene. 
The higher precision score means that there are more actual graphene pixels among the pixels predicted as graphene by our model. If the precision is lower, the portion of background pixels classified as graphene is higher. 
\begin{equation}
\textrm{Precision} = \frac{\textrm{TP}} {\textrm{TP + FP}}
\label{eqPrecision}
\end{equation}
Similarly, we define the recall in Eq.~(\ref{eqRecall}). 
If the recall value is high, a graphene pixel is less likely to be classified as a background pixel. 
\begin{equation}
\textrm{Recall} = \frac{\textrm{TP}}{\textrm{TP + FN}}
\label{eqRecall}
\end{equation}
Lastly, the F1 score in the Eq.~(\ref{eqF1}) is nothing more than the harmonic mean between the precision and the recall. 
\begin{equation}
\textrm{F1 score} = \frac{2 \times \textrm{Precision} \times \textrm{Recall}}{\textrm{Precision + Recall}}
\label{eqF1}
\end{equation}

The evaluation of the four tree-based classification algorithms using the four metrics are shown in Table \ref{table_metric}. 
The accuracies for the four algorithms are more or less the same but DT has the smallest accuracy compared to others. DT also has far less precision around $\sim 74 \%$ in contrast with the other methods which are above $\sim 88 \%$. LGBM has the highest precision $\sim 91\%$. Regarding the recall values, on the other hand, DT scores above $\sim 78 \%$ which is superior to others. 
The RF, XGB, and LGBM record around $\sim 81-82\%$ in F1 score, while DT is lower by around $6 \%$ with respect to the other methods.

\begin{figure}[t]
%\begin{figure*}[t]
\begin{center}
\includegraphics[width=0.45\textwidth]{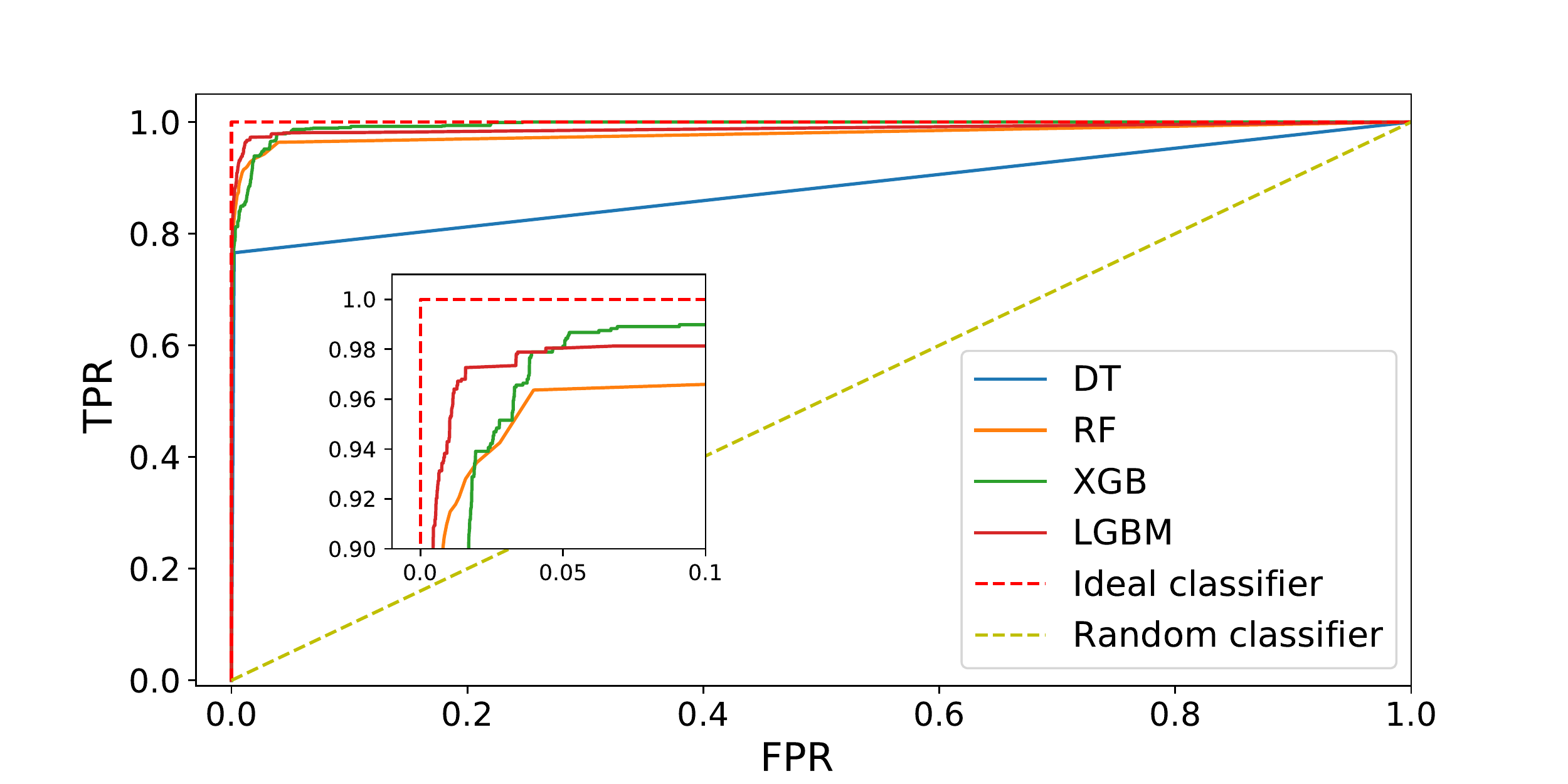}
\end{center}
\caption{The receiver operating characteristic (ROC) curves for random classifier (yellow dashed), ideal classifier (red dashed), DT (blue solid), RF (orange solid), XGB (green solid), and LGBM (pink solid). The inset shows the same ROC curve for the finer range of false positive range (FPR) in Eq.~(\ref{eqFPR}) near 0.0 and the true positive range (TPR) near 1.0.}
\label{ROC_CURVE}
\end{figure}

We further employ several additional performance measures such as the receiver operating characteristic (ROC) curves, their area under the curve (AUC), and the intersection over union (IOU). We show the ROC curve in Fig.~\ref{ROC_CURVE} as a function of the true positive rate (TPR) which stands for the recall, and the false positive rate (FPR) which
%also means $(1-\textrm{Sensitivity})$ and 
is defined as
\begin{equation}
\textrm{False Positive Rate (FPR)} = \frac{\textrm{FP}}{\textrm{FP + FN}}.
\label{eqFPR}
\end{equation}

We get these ROC curves by changing the criteria of rounding off the probability between $0$ (background) and $1$ (graphene) for each pixel obtained from each ML model, leading to different values of TPR and FPR. The AUC for the ideal classifier whose ROC is denoted by the red dashed line is $1$ using normalized units, whereas the AUC for the random classifier denoted by the yellow dashed line is $1/2$ as shown in Fig.~\ref{ROC_CURVE}. The AUCs for the four different classification algorithms which range between $1/2$ and $1$ are presented in Table.~\ref{table_AUC}. 
Their AUCs have an order such that DT $\ll$ RF $<$ LGBM $<$ XGB. One can see in the inset of Fig.~\ref{ROC_CURVE} in details that XGB has a smaller TPR than RF and LGBM at nearly zero FPR but it starts to surpass the others when FPR $\gtrsim 0.05$.

\begin{table}[]
\begin{tabular}{|c|c|}
\hline
 Algorithm     & AUC    \\ \hline
DT & 0.8731 \\ \hline
RF & 0.9792 \\ \hline
XGB       & 0.9942 \\ \hline
LGBM      & 0.9884 \\ \hline
\end{tabular}
\caption{Area under the curve (AUC) score for the four tree-based ML algorithms such as decision tree (DT), random forest (RF), XGBoost (XGB), and LightGBM (LGBM).}
\label{table_AUC}
\end{table}

To evaluate the performance of the pixel-wise segmentation method we use the intersection over union (IOU) index  defined as % Eq.~(\ref{eqIOU}). 
\begin{equation}
\textrm{IOU} = \frac{\textrm{Target Pixel} \cap \textrm{Predicted Pixel}}{\textrm{Target Pixel} \cup \textrm{Predicted Pixel}}.
\label{eqIOU}
\end{equation}
This index indicates how many pixels overlap between the target pixels and 
the predicted pixels and returns a value in the range between $0$ and $1$. 
The IOU values for the four algorithms are presented in 
Table~\ref{table_IOU} and they have an order of DT $\ll$ RF $\ll$ 
LGBM $<$ XGB which indicates that the XGB is the most suitable algorithm 
for pixel-wise segmentation of our datasets.

The RF classifier  has the highest accuracy and F1 score, but its pixel-resolved segmentation shows a relatively low IOU score that is slightly higher than the DT. On the other hand, the XGB has the highest IOU score but less accuracy, precision, recall, and F1 score than RF. The LGBM has higher precision and IOU score than RF. 
In practice we can choose the classifier depending on which metric or index is more relevant for the classification task at hand or mix more than one classifier as we will explain in the next section.

\begin{table}[]
\begin{tabular}{|c|c|}
\hline
Algorithm    & IOU Score (\%) \\ \hline
DT & 54.61         \\ \hline
RF & 63.86         \\ \hline
XGB      & 71.18         \\ \hline
LGBM      & 70.73         \\ \hline
\end{tabular}
\caption{Evaluation of the intersection over union (IOU) score for the four tree-based ML algorithms such as decision tree (DT), random forest (RF), XGBoost (XGB), and LightGBM (LGBM).}
\label{table_IOU}
\end{table}

\begin{table}[]
\begin{tabular}{|c|c|}
\hline
Algorithm    & Inference time (s) \\ \hline
DT & 0.1185         \\ \hline
RF & 0.1413         \\ \hline
XGB      & 0.2822         \\ \hline
LGBM      & 0.2006       \\ \hline
\end{tabular}
\caption{The inference time per image for the four tree-based ML algorithms such as decision tree (DT), random forest (RF), XGBoost (XGB), and LightGBM (LGBM).}
\label{table_time}
\end{table}

Regarding the inference time, as shown in Table.~\ref{table_time}, the DT algorithm is the fastest model with a computation time of $\sim$0.1 seconds per image, while the XGB is the slowest one at $\sim$0.3 seconds per image using Intel(R) Core(TM) i5-10400 CPU. In practice, this means that we can analyze a thosand images in a matter of minutes regardless of the method used.

%\end{figure*}

%
\begin{table*}[]
\begin{tabular}{|c|c|c|c|c|c|}
\hline
\diaghead(-2,1){aaaaaaaaaaaaaa}%
{Algorithm}{metric}                                               & Accuracy (\%) & Precision (\%) & Recall (\%) & F1 score (\%) & IOU score (\%) \\ \hline
\begin{tabular}[c]{@{}c@{}}RF+XGB\\ (precision, recall, F1 score) \end{tabular}  & 99.77        & 89.36         & 72.95       & 79.96        & 66.82        \\ \hline
\begin{tabular}[c]{@{}c@{}}RF+LGBM \\ (precision) \end{tabular}  & 99.78        & 91.23         & 73.37       & 80.96        & 66.82          \\ \hline
\begin{tabular}[c]{@{}c@{}}RF+LGBM\\(recall, F1 score) \end{tabular}  & 99.77        & 89.36         & 72.95       & 79.96        & 66.82          \\ \hline
\begin{tabular}[c]{@{}c@{}}XGB+LGBM \\ (precision, F1 score) \end{tabular} & 99.78        & 91.23         & 73.37       & 80.96        & 68.22         \\ \hline
\begin{tabular}[c]{@{}c@{}}XGB+LGBM \\ (recall) \end{tabular} & 99.77        & 88.01         & 75.26       & 80.80        & 67.98         \\ \hline
\begin{tabular}[c]{@{}c@{}}RF+XGB+LGBM\\ (precision, recall, F1 score) \end{tabular} & 99.78        & 90.64         & 73.82       & 81.03    &   68.28      \\ \hline
\end{tabular}
\caption{Algorithm choice versus metric table for five different combinations of the three single classification algorithms random forest (RF), XGBoost (XGB), and lightGBM (LGBM) using the five metrics accuracy, precision, recall, F1 score, and IOU score. 
%{\bf We specify within parentheses the metrics used to optimize the weights if the results are different depending on the metrics.}
In parentheses below each one of the fused-classifiers, we specify the metrics used to optimize the weights of the classifiers. We listed several metrics when their performances were the same down to two decimal places. 
}

\label{table_fuse}
\end{table*}

\subsection{Fusions between classifiers}
In this section, we combine the different single classifiers RF, XGB, and LGBM in an attempt to improve the performance of our model as proposed in Ref.~\cite{li2014classifier} that calculates different weights for each classifier depending on metrics accuracy. We will show that it is possible to improve the overall performance of different metrics of our choice at the expense of slightly reducing the IOU. We exclude the DT as it scores far fewer points particularly in the precision, F1 score, AUC and IOU. We predict to find a monolayer graphene pixel with the probability $P$ defined in Eq.~(\ref{eqfusion}) for each pixel in an OM image when we utilize the combined probability consisting of $N$ single classifiers through
\begin{equation}
%P = w_1*P_1+w_2*P_2+w_3*P_3 
P = \sum_i^N w_i P_i,
\label{eqfusion}
\end{equation}
where $P_i$ stands for the probability of finding a monolayer graphene at that pixel using the $i^{\rm th}$ single classifier, and $w_i$ is the $i^{\rm th}$ classifier’s weight which is given as \cite{li2014classifier}
\begin{equation}
w_i = \frac{{\rm metric}(i)^{10}}{\sum_{j=1}^{N} {\rm metric}(j)^{10}},
\label{eqweight}
\end{equation}
where metric($i$) represents the $i^{th}$ single classifier’s metric. In other words, the probability $P$ of a joint classification model is defined such that the probabilities of finding a monolayer graphene at that pixel using single classifiers are
expressed as a linear combination of each classifier's probability multiplied by the respective weight coefficients. 
Like in the single classifiers, we round off the final probability to either 0 (background) and 1 (graphene). Then we use $P$ at each pixel to calculate the metrics as presented in Table.~\ref{table_fuse}. We take into account four combinations, RF+XGB, RF+LGBM, XGB+LGBM, RF+XGB+LGBM. We calculate the weights $w_i$ for the $i^{\rm th}$ single classifier as defined in Eq.~(\ref{eqweight}) taking one of the four metrics -- accuracy, precision, recall, and F1 score. We indicated within parentheses the reference metric used to calculate the weights of the classifiers, and listed more than one when they gave the same results down to two decimal places. We exclude the accuracy as a weight because the accuracies of RF, XGB, and LGBM are all on the order of $\sim 99.7 \%$.

The single classifier RF has a higher accuracy of $99.79 \%$, recall of $76.54 \%$, and F1 score of $ 82.09\%$, but it has a smaller IOU score of $63.86\%$ than XGB and LGBM. If we join XGB with RF, regardless of which metric we use, we get an improved IOU score by $\sim 3 \%$, but sacrificing accuracy by $0.01 \%$, precision by $0.25 \%$, recall by $\sim 3.6 \%$, and the F1 score by $\sim 2.1 \%$ compared to using the single RF classifier. On the other hand, we get improved accuracy by $0.05 \%$, precision by $\sim 1.4 \%$, sacrificing recall by $2.3\%$, F1 score by $\sim 0.8\%$, and the IOU score by $\sim 4.4\%$ compared to using the single XGB classifier. 

If we join LGBM with RF, the results are different depending on the metrics we use for a weight. In the case of precision as a weight, it has $\sim 1.6 \%$ higher precision, $\sim 3\%$ higher IOU score than a single RF classifier. If we use either recall or F1 score as a metric, it has benefit in IOU score by $\sim 3\%$ compared to the single RF. On the other hand, combining LGBM with RF has no merit in comparison with the single LGBM. 

If we combine XGB and LGBM, in the case where either precision or the F1 score is chosen as a weight, the accuracy is improved to $99.78 \%$, and the IOU hits $68.22 \%$ which is less than both XGB ($71.18 \%$) and LGBM ($70.73 \%$) separately. The precision ($91.23 \%$), recall ($73.37 \%$), and F1 score ($80.96 \%$) are the same as those of a single classifier of LGBM. When we use the recall as a weight, the IOU hits $67.98 \%$ which is again smaller than that of single XGB and LGBM classifiers. The accuracy is the same as that of LGBM ($99.77 \%$), but the precision ($88.01 \%$), recall ($75.26 \%$), and F1 score ($80.80 \%$) have the same value as those of XGB. 

Lastly, if we combine all three classifiers, all metrics and indices are overall averaged regardless of the metrics used for the weights. The accuracy and F1 score are improved in comparison with single XGB and LGBM classifiers, and the precision becomes higher than single RF and XGB. The recall becomes higher than a single LGBM, and the IOU is improved compared to a single RF.

The resulting fused-classifier metrics presented in Table~\ref{table_fuse} indicates that we can indeed improve the overall performance of our models if we target a particular metric to optimize, and at times it is possible to achieve an overall improvement in several metrics at the expense of a small decrease in the IOU by a few percents with respect to the XGB and LGBM single classifier scores.

\section{Conclusions}\label{sec5}
The preparation of graphene electronic devices during almost the last two decades has relied on human inspection during many hours of routine scanning in the entire space of samples to identify the number of graphene layers deposited on the substrate. 
Recently several research teams have attempted to introduce ML in this process to reduce such time-consuming and error-prone human labor. However, the existing proposals are mostly focused on developing DNN or CNN based models which require quite a large number of labeled training dataset, and this threshold makes practical application difficult. In this paper, we have proposed a pixel-wise tree-based ML classification tool that can be used for samples prepared under consistent scanning conditions such as illumination or the thickness of the substrate. Our method can distinguish a specific number of graphene layers from the background and performs precisely even with a small number of training images, in contrast to existing works utilizing DL tools that require hundreds to thousands of image data to train a model. 

We have trained our ML model with four different tree-based algorithms such as DT, RF, XGB, and LGBM, and examined their outcomes using several metrics and indices -- accuracy, precision, recall, F1 score, ROC, AUC, and IOU. There is no absolute best model among the single classifiers as they have different strengths and weaknesses in metrics and indices, and those metrics vary depending on the data that we deal with. We have shown that it is possible to improve the performance by combine more than one classifier.
Considering overall performance, we propose that the LGBM or XGB are good choices when using a single classifier, and for the fused classifiers model, the RF+XGB+LGBM shows satisfactory performance and can be chosen in routine applications. 

In this work, we have sorted out the monolayer graphene from the OM images, but the same process can be performed to find any other number of multilayer graphene or for any other 2D vdW materials such as hexagonal boron nitride, transition metal chalcogenides or black phosphorus. 
Furthermore, our model requires only a few images and costs less computational time than a few minutes in total to train a ML model even without using a GPU. Therefore, we expect our work will be of immediate utility for researches on 2D materials, and will greatly ease the repetitive and time-consuming tasks in experiments.

\begin{acknowledgments}
This work was supported by Korean NRF through the Grants No. 2020R1A5A1016518 (W.H.C), No. 2021R1A6A3A01087281 (J.S.), No. 2020R1A2C3009142 (J.J.).
YDK was supported by the KIST Institutional Program (Project  No.2E31781-22-108).
We acknowledge computational support from KISTI Grant No. KSC-2021-CRE-0389 and by the computing resources of Urban Big data and AI Institute (UBAI) at UOS and the network support from KREONET. 
We also acknowledge partial support for W.H.C. from the Korean Ministry of Land, Infrastructure and Transport (MOLIT) from the Innovative Talent Education Program for Smart Cities.
\end{acknowledgments}

\bibliographystyle{unsrt}
\bibliography{woon_draft_revised.bib}

\end{document}